\documentclass[preprint,showpacs,preprintnumbers,amsmath,amssymb, superscriptaddress,longbibliography,nofootinbib,showkeys]{revtex4-1}

\usepackage{textcomp}
\usepackage{makeidx}
\usepackage{amsmath}
\usepackage{subfig}
\usepackage{amssymb}
\usepackage{hyperref}
\usepackage{graphicx}
\usepackage{epstopdf}
\usepackage{color}
\usepackage{soul}

\begin{document}

\title{Gravitational Decoupling algorithm modifies the value of the conserved charges and thermodynamics properties in Lovelock Unique Vacuum theory.}

\author{Milko Estrada}
\email{milko.estrada@ua.cl}

\affiliation{Departmento de f\'isica, Universidad de Antofagasta, 1240000 Antofagasta, Chile}

\date{\today}


\begin{abstract}
We provide an extension of the Gravitational Decoupling (algorithm) for the Lovelock theory with Unique Vacuum (LUV), which represents a simple way to solve the equations of motion. Due to the application of this algorithm, the energy of the system splits in the {\it energy of the seed solution} and the {\it energy of each quasi-LUV sector}. Under certain assumptions imposed, the total energy varies due to the contribution of energy of each quasi-LUV sector. It is provided a new solution, whose energy differs from the energy of the seed solution in a quantity that depends on the number of extra sources. The new solution has two inner horizons, which is a proper characteristic of itself. Furthermore, its thermodynamics differs from the seed solution, since our solution is always stable and  does not have phase transitions. Since the first law of thermodynamics is modified by the presence of the matter fields, we provide a new version of
the first law for LUV, where a local definition of the variation of energy is defined, and, where the entropy
and temperature are consistent for LUV theory.

\end{abstract}

\maketitle

\section{Introduction}
Finding new solutions of physical interest to the Einstein field equations is not an easy task due to the highly nonlinear behavior of its equations of motion. In this connection, in (2017) it was proposed the  Gravitational Decoupling method (GD) \cite{Ovalle:2017fgl}, which represents an easy algorithm to decouple gravitational sources in General Relativity. One interesting extension of this method in reference \cite{Ovalle:2019qyi}. This algorithm involves a Minimal Geometric Deformation (MGD) to the metric tensor of one seed solution. Furthermore, the sources are decoupled in one seed and one extra source. The method was described in reference \cite{Ovalle:2017wqi} as follows: ``given two gravitational sources: a seed source A and an extra source B,  standard Einstein’s equation are first solved for A, and then a simpler set of  quasi-Einstein equations are solved for B. Finally, the two solutions can be combined in order to derive the complete solution for the total system". Since its appearance, by applying the algorithm to previously known seed solutions, this method has served to find several new solutions of physical interest for GR, as for example stellar distributions \cite{Ovalle:2017wqi,Maurya:2020rny,Tello-Ortiz:2020ydf,Maurya:2019noq,Torres:2019mee,Hensh:2019rtb,Ovalle:2019lbs,Maurya:2019wsk,Estrada:2018vrl,Morales:2018urp,Morales:2018nmq} and black hole solutions \cite{Ovalle:2018umz,Contreras:2021yxe,Ovalle:2020kpd,Rincon:2020izv,Tello-Ortiz:2020ydf,Contreras:2018nfg}. See other applications in references \cite{daRocha:2020gee,daRocha:2020jdj,Arias:2020hwz,Casadio:2019usg,Estrada:2019aeh,Gabbanelli:2019txr,daRocha:2019pla,Contreras:2019iwm,Fernandes-Silva:2019fez,Sharif:2018tiz,Panotopoulos:2018law,Sharif:2018toc,Heras:2018cpz,Estrada:2018zbh,Gabbanelli:2018bhs,Estrada:2020ptc}.

The recent detection of gravitational waves through the collision of two rotating black holes \cite{Abbott:2016blz,Abbott:2017oio}, together with the also recent assignment of the Nobel Prize, 
 have positioned to the black holes as one of the most interesting and intriguing objects in gravitation. In this connection, the fact that the black holes, due to quantum fluctuations, emit as black bodies, where its temperature is related to its surface gravity \cite{Hawking:1974sw}, shows that in these objects the geometry and thermodynamics are directly connected. 
 
 Applying GD method to the Schwarzschild solution, new black hole solutions have been found. As for example, three interesting solutions, characterized by  isotropic, conformal and barotropic equations of state, were found in reference \cite{Ovalle:2018umz}. Other interesting solutions with hair were found in reference \cite{Ovalle:2020kpd}. Non-trivial extensions of the Kerr and Kerr-Newman
black holes with primary hair were found in reference \cite{Contreras:2021yxe}. Thus, it is undoubtedly interesting the search of new black hole solutions by using the GD method and test its physical behaviors.

On the other hand, during the last years, several branches of theoretical physics have predicted the existence of extra dimensions. Now, considering higher dimension in gravity opens up a range of new possibilities that retain the core of the Einstein gravity in four dimensions. Lovelock theory \cite{Lovelock:1971yv} is one of these possibilities, as, although includes higher powers of curvature corrections in the action, its equations of motion are of second order on the metric and the energy momentum tensor is conserved.
It is worth to mention that the higher curvature terms can modify the structure of the black hole solutions, showing different properties with respect to the solutions in General Relativity. So, the study of BH solutions with presence of higher curvature terms is undoubtedly interesting from the physical point of view. One interesting extension of the Gravitational Decoupling algorithm for Pure Lovelock gravity \cite{Cai:2006pq} was developed in reference \cite{Estrada:2019aeh}, showing a regular black hole solution. Some examples of recent studies about black holes in Lovelock gravity in references \cite{Konoplya:2020kqb,Barton:2021wfj,Fan:2019aoj,Xu:2019krv,Fairoos:2018pee}. 

One potential drawback of the generic Lovelock gravity is the existence of more than a single {\it ground state}, namely more than a single constant curvature spaces solution, or equivalently, more than a single potential effective cosmological constants \cite{Camanho:2011rj}. The potential effective cosmological constants can be complex numbers,
which makes those ground states unstable under dynamical evolution. One way to avoid this latter is to choose the coupling constant $\{\gamma_p\}$ such that the equations of motion have, roughly speaking, the form $(R-\Lambda)^{n} = 0$. In this case there is a single, but $n$-fold degenerated ground state of constant curvature. This case is known as {\it Lovelock with unique vacuum} (LUV) or  Lovelock with $n$ fold degenerated ground state.  The vacuum and static solution with an AdS ground state was studied in references \cite{Crisostomo:2000bb,Aros:2000ij} and with a dS ground state in reference \cite{Aros:2008ef}. Regular black hole solutions for this theory have been studied in references \cite{Aros:2019quj,Aros:2019auf}.

On the other hand, in (2020), has appeared a new definition of conserved charges in refereces \cite{Aoki:2020prb,Sorge:2020pdj}. In \cite{Aoki:2020prb} the energy and momentum can be computed by integrating a covariantly conserved current $J^\mu=T^\mu_\nu \xi^\nu$ in a volume integral. It is worth to mention that the definition of reference \cite{Aoki:2020prb} can be reduced to the definition of conserved energy of reference \cite{Sorge:2020pdj} for a Killing vector $\xi^\mu=-\delta^\mu_0$. 
It is worth to mention that this definition is diffeo-invariant. Furthermore the current is covariantly conserved due that $\nabla_\mu T^\mu_\nu=0$ and $\nabla_\mu \xi_\nu+\nabla_\nu \xi_\mu=0$. This definition differs from Komar charge \cite{Kastor:2008xb} in the fact that this latter diverges at infinity for a non zero cosmological constant. This divergence can be suppressed by inserting counter terms in the action \cite{Aros:1999kt}. Furthermore, the Komar charge vanishes for $(2+1)$ dimensions. In this respect, the definition of \cite{Aoki:2020prb} has the adventages that does not diverges at infinity for $\Lambda \neq 0$, does not vanishes for $(2+1)$ dimensions, and represents a simple mathematical tool without the need of included counter terms in the action. 

In a genuine way, in reference \cite{Estrada:2020ptc} was discovered that, using the definition of conserved charges of reference \cite{Aros:1999kt} ( which considers counter terms), the mass of a particular type of regular black hole increases due to the application of the gravitational decoupling algorithm for the General Relativity. So, due to the adventages above mentioned of the definition of conserved charges of reference \cite{Aoki:2020prb}, it is of physical interest to test how the mass of black holes varies, using this definition by applying the Gravitational Decoupling algorithm for Lovelock Gravity With Unique Vacuum. 

In this work we will provide an extension of the Gravitational Decoupling algorithm for Lovelock with Unique Vacuum theory. We will test the effects of the GD algorithm on the value of the conserved charges of a black hole. Particularly, we will test how the addition of extra sources modify the value of the energy of a black hole. We will show a new analytical black hole solution and test its variation of energy, due to the application of the GD algorithm. Furthermore, we will compare the horizons structure and the thermodynamics behavior of the obtained solution with the solutions previously studied in the literature. On the other hand, since the first law of thermodynamics is modified by the presence of matter fields in the energy momentum tensor \cite{Ma:2014qma,Estrada:2020tbz}, we will provide a new version of
the first law for LUV, it order to obtain the correct values of entropy and temperature for LUV theory.

\section{Generic Lovelock theory and the case with unique vacuum}

The generic Lovelock Lagrangian is :
\begin{equation}\label{LovelockLagrangian}
L  = \sqrt{-g} \sum_{p=0}^n \gamma_p L_p,
\end{equation}
where $n=\frac{d}{2}-1$  for $d$ even and $n=\frac{d-1}{2}$ for $d$ odd and, $\gamma_p$ are arbitrary coupling constants. $L_p$ is a topological density defined as:
\begin{equation}
 L_p = \frac{1}{2^p} \delta^{\mu_1 \nu_1 ...\mu_p \nu_p}_{\alpha_1 \beta_1 ... \alpha_p \beta_p} \displaystyle \Pi^p_{r=1} R^{\alpha_r \beta_r}_{\mu_r \nu_r},
\end{equation}
where $R^{\alpha \beta}_{\mu \nu}$ is a $p$ order generalization of the Riemann tensor for the Lovelock theory, and:
\begin{equation}
 \delta^{\mu_1 \nu_1 ...\mu_p \nu_p}_{\alpha_1 \beta_1 ... \alpha_p \beta_p} = \frac{1}{p!} \delta^{\mu_1}_{\left[\alpha_1\right.} \delta^{\nu_1}_{\beta_1} ... \delta^{\mu_p}_{\alpha_p} \delta^{\nu_p}_{\left.\beta_p \right]}   
\end{equation}
is the {\it generalized Kronecker delta}.

It is worth to stress that, the terms $L_0,L_1$ and $L_2$ are proportional to the cosmological constant, Ricci Scalar and the Gauss Bonnet Lagrangian, respectively. The corresponding equation of motion is given by: 
\begin{equation}
 \sum ^n_{p=0} \gamma_p \mathcal{G}^{(p)}_{AB} =   T_{AB},
\end{equation}
where $\mathcal{G}^{(p)}_{AB}$ is a $p$ order generalization of the Einstein tensor due to the topological density $L_p$. As example $\mathcal{G}^{(1)}_{AB}$  is just the Einstein tensor, $G_{A B}$, associated with the Ricci scalar and $\mathcal{G}^{(2)}_{AB}$ is the Lanczos tensor, $H_{A B}$, associated with the Gauss Bonnet Lagrangian.

As for example, the Einstein Gauss Bonnet equations of motion up to $n=2$, with cosmological constant are:

\begin{equation}
\gamma_1 G^{A}_{B} +\gamma_0  \Lambda \delta^A_B +\gamma_2 H^{A}_{B} = T^{A}_{B}  
\end{equation}
where the Lanczos tensor is:
\begin{equation}
H_{AB} = 2\Big (R R_{AB} - 2R_{AC}R^C_B - 2R^{CD}R_{ACBD}  + R^{CDE}_{A} R_{BCDE} \Big)- \frac{1}{2} g_{AB} L_2.
\end{equation}

\subsection{Lovelock with unique vacuum}
As mentioned in the introduction, Lovelock theory can be factorized in several effective cosmological constants \cite{Camanho:2011rj}. In this connection, for $\gamma_p=0$ from $p > I$, the vacuum equations of motion can be written as \cite{Aros:2019quj}:

\begin{equation}\label{LovelockLangrangianalternative}
 G^{\mu}_{ (LL)\hspace{1ex} \nu} \propto \delta_{\mu_1 \nu_1 \ldots \mu_{I}\nu_I \nu}^{\alpha_1\beta_1 \ldots \alpha_{I}\beta_I\mu} (R^{\nu_1 \mu_1}_{\hspace{2ex}\alpha_1 \beta_1} + \kappa_1 \delta^{\mu_1 \nu_1}_{\hspace{2ex}\alpha_1 \beta_1})\ldots (R^{\nu_I \mu_I}_{\hspace{2ex}\alpha_I \beta_I} + \kappa_I \delta^{\mu_I \nu_I}_{\hspace{2ex}\alpha_I \alpha_I}) =0. 
\end{equation}

This shows, as expected, that the Lovelock gravity can be factorized in several ground states of constant curvature. To analyze those backgrounds one can introduce the ansatz $ R^{\nu_1 \mu_1}_{\hspace{2ex}\alpha_1 \beta_1}  = x \delta^{\mu_1 \nu_1}_{\hspace{2ex}\alpha_1 \beta_1}$. This maps \eqref{LovelockLangrangianalternative} into $G^{\mu}_{\nu} = P_{l}(x) \delta^{\mu}_{\nu}$ where

\begin{equation}\label{IndexialEq}
  P_{l}(x) = \sum_{p=0}^I \gamma_p x^p = (x+\kappa_I)\ldots(x+\kappa_1).
\end{equation}
where $\kappa_i$ must be a real number 
 $ \forall \gamma_p \in \mathbb{R}$
 
One way of avoiding equations of motion with several ground states is consider the following action:
\begin{equation} \label{AccionBHS}
    S_n= \int \displaystyle \sum_{p=0}^{p=n} \gamma^n_p L_p.
\end{equation}

It is worth to mention that for $n=1$ the Einstein Hilbert action is recovered \cite{Crisostomo:2000bb}.

In order to obtain $\kappa_1=\kappa_2=...=\kappa_I$, {\it i.e.} a $n$-fold degenerated vacuum state, we choose the following coupling constants \cite{Crisostomo:2000bb}

\begin{equation} \label{ConstantesAcoplamientoBHS}
\gamma^n_p = \left\{\begin{array}{cl}
                    \frac{l^{2(p-n)}}{d-2p}\binom{n}{p} &\textrm{ for } 0 \leq p \leq n\\
                    0 &\textrm{ for } n < p 
                  \end{array}
\right.
\end{equation}
So, in the references \cite{Crisostomo:2000bb,Aros:2000ij} it was shown that, using this coupling constants, the equations of motion adopt the following form \cite{Aros:2019quj}:
\begin{equation}\label{EcuacionMovimientoBHS}
  \frac{\delta }{\delta g_{\mu\nu}} L\sqrt{g} \sim ((R \pm l^{-2})^n)^{\mu\nu}=0.
\end{equation}

So, the Lovelock theory has a unique vacuum (A)dS, but $n$-fold degenerated.

\section{ LUV equations of motion for multiples sources} \label{seccion2}

In this work we study the static $d$ dimensional spherically symmetric metric:

\begin{equation}
ds^2=-\mu(r)dt^2+\frac{dr^2}{\mu(r)}+r^2 d\Omega^2_{d-2}, \label{metrica1}
\end{equation}
where $d\Omega^2_{d-2}$ corresponds to the metric of a $(d-2)$ unitary sphere. The energy momentum tensor corresponds to a  neutral perfect fluid:

\begin{equation}
T^A_B=\mbox{diag}(-\rho,p_r,p_\theta,p_\theta,...),    
\end{equation}
where, from the spherical symmetry, we have for all the $(d-2)$ angular coordinates that $p_\theta=p_\phi=...$. It is worth to mention that this form of metric \eqref{metrica1} imposes that $T^0_0=T^1_1$ {\it i.e.} $-\rho=p_r$. 

For the cosmological constant:
\begin{equation} \label{ConstanteCosmologica}
    \Lambda =\mp \frac{(d-1)(d-2)}{2l^2},
\end{equation}
the equations of motion are \cite{Crisostomo:2000bb,Aros:2019quj}:
\begin{equation} \label{EqMotion}
\frac{d}{dr} \left (  r^{d-1} \left [ \frac{1-\mu(r)}{r^2} \pm \frac{1}{l^2}  \right ]^n  \right ) = r^{d-2} \rho(r)
\end{equation}

with $d-2n-1 \ge 0$. It is easy to check that the equation \eqref{EqMotion} takes the positive (negative) branch for negative (positive) cosmological constant. The conservation law $T^{AB}_{;B}=0$ gives:
\begin{equation}
p'_r+\frac{d-2}{r}(p_r-p_\theta ) =0 . \label{conservacion1}
\end{equation}

We start by decoupling the energy momentum tensor into a seed energy momentum tensor, $\bar{T}^A_B$, and $n$ extra sources
\begin{equation} \label{EM1}
    T^A_B=\bar{T}^A_B+ \alpha (\theta_1)^A_B + \alpha^2 (\theta_2)^A_B+...+\alpha^{n-1}(\theta_{n-1})^A_B+\alpha^n (\theta_{n})^A_B
\end{equation}
therefore the number of sources is determined by the value of $n$, {\it i.e} depend on the power of the Riemann tensor. 
The seed energy momentum tensor has the form $\bar{T}^A_B=\mbox{diag}(-\bar{\rho},\bar{p}_r,\bar{p}_\theta,\bar{p}_\theta,...)$. So, it is easy to check that:
\begin{align}
    \rho=&\bar{\rho}- \alpha (\theta_1)^0_0- \alpha^2 (\theta_2)^0_0 -...  - \alpha^{n-1} (\theta_{n-1})^0_0 - \alpha^n (\theta_n)^0_0            \label{densidadefectiva} 
\end{align}
\begin{align}
    p_r =& \bar{p}_r + \alpha (\theta_1)^1_1 + \alpha^2 (\theta_2)^1_1 +... + \alpha^{n-1} (\theta_{n-1})^1_1 + \alpha^n (\theta_n)^1_1 \label{pradialefectiva} 
\end{align}
\begin{align} 
    p_\theta =&  \bar{p}_{\theta} + \alpha (\theta_1)^2_2 + \alpha^2 (\theta_2)^2_2+...+ \alpha^{n-1} (\theta_{n-1})^2_2 + \alpha^n (\theta_n)^2_2 \label{ptangencialefectiva}
\end{align}

Due that the form of metric imposes that $-\rho=p_r$, we impose arbitrarily and for convenience that $-\bar{\rho}=\bar{p}_r$ and $(\theta_i)^0_0=(\theta_i)^1_1$. On the other hand, due that $p_\theta=p_\phi=...$, we impose arbitrarily that $\bar{p}_{\theta}=\bar{p}_{\phi}=...$ and $(\theta_i)^2_2=(\theta_i)^3_3=...$.

Thus, replacing equation \eqref{densidadefectiva} into the equation \eqref{EqMotion}, we obtain the $(t,t)$ and $(r,r)$
components of the equations of motion:

\begin{align} \label{EqMotion1}
\frac{d}{dr} \bigg (  r^{d-1} \bigg [ \frac{1-\mu(r)}{r^2} \pm \frac{1}{l^2}  \bigg ]^{n}  \bigg ) &= r^{d-2} \bigg ( \bar{\rho}- \alpha (\theta_1)^0_0- \alpha^2 (\theta_2)^0_0 -... \nonumber \\
&- \alpha^{n-1} (\theta_{n-1})^0_0 - \alpha^n (\theta_n)^0_0   \bigg)
\end{align}
 
We solve the $(t,t)$ and $(r,r)$ components of the LUV equations together with the conservation equation. Using  the Bianchi identities, we ignore the remaining $(\theta,\theta)=(\phi,\phi)=...$ components (the suspense points indicate that all the tangential components of the LUV equations are similar).

By inserting equations \eqref{densidadefectiva},\eqref{pradialefectiva} and \eqref{ptangencialefectiva} into equation \eqref{conservacion1}:

\begin{align} \label{conservacion3}
\bar{p}_r'+ \frac{d-2}{r}(\bar{p}_r-\bar{p}_t ) &+  \alpha \Big ( \big ((\theta_1)^1_1 \big )' +\frac{d-2}{r} \big ((\theta_1)^1_1-(\theta_1)^2_2 \big ) \Big ) \nonumber \\
&+\alpha^2 \Big ( \big ((\theta_2)^1_1 \big )' +\frac{d-2}{r} \big ((\theta_2)^1_1-(\theta_2)^2_2 \big ) \Big ) +...\nonumber \\
&+\alpha^{n-1} \Big ( \big ((\theta_{n-1})^1_1 \big )' +\frac{d-2}{r} \big ((\theta_{n-1})^1_1-(\theta_{n-1})^2_2 \big ) \Big )  \nonumber \\
&+\alpha^n \Big ( \big ((\theta_n)^1_1 \big )' +\frac{d-2}{r} \big ((\theta_n)^1_1-(\theta_n)^2_2 \big ) \Big )=0
\end{align}

Thus, the system to solve corresponds to equations \eqref{EqMotion1} and \eqref{conservacion3}. 

The LUV equations of motion for the seed energy momentum tensor are recovered for the limit $\alpha \to 0$:
\begin{equation} \label{conservacionImpuesta1}
\nabla_A \bar{T}^A_B=0,
\end{equation}

 For $n=1$ both components of energy momentum tensor are directly conserved, {\i.e} $\nabla_A(\theta_1)^A_B=0$. However, for $n>1$ one can notice that:
\begin{equation}
\alpha \nabla_A(\theta_1)^A_B + \alpha^2 \nabla_A(\theta_2)^A_B+...+\alpha^{n-1} \nabla_A(\theta_{n-1})^A_B 
+\alpha^n \nabla_A(\theta_{n})^A_B=0,
\end{equation}
where the covariant derivative is computed by using the line element \eqref{metrica1}.  In this work we impose in arbitrarily way that:
\begin{equation} \label{conservacionImpuesta2}
 \alpha^i \nabla_A(\theta_i)^A_B=0,
   \end{equation}
 with $\alpha^i \neq 0$. Thus, we will solve the system \eqref{EqMotion1}, \eqref{conservacionImpuesta1} and \eqref{conservacionImpuesta2}. Under this assumption each source is separately conserved, and thus, there is no exchange of energy momentum between them. 
Therefore, our energy momentum tensor \eqref{EM1} is a specific way of decoupling the system inspired by the approach of reference \cite{Ovalle:2017fgl}.

\section{ Gravitational decoupling by MGD for LUV} \label{metodo}
  
We start with a solution to the equations \eqref{EqMotion1}, \eqref{conservacionImpuesta1} and \eqref{conservacionImpuesta2} with $\alpha=0$, namely {\it seed solution} $\{\bar{\mu},\bar{\rho},\bar{p}_r,\bar{p}_t \}$, where $\bar{\mu}$ is the corresponding metric function:

\begin{equation}
    ds^2=-\bar{\mu}(r) dt^2+\bar{\mu}(r)^{-1} dr^2+r^2 d\Omega^2_{d-2}. \label{metrica2}
\end{equation}

Turning on the parameter $\alpha$, the effects of the sources $(\theta_i)_{A B}$ appear on the seed solution. These effects can be encoded in the geometric deformation undergone by the seed fluid geometry $\{\bar{\mu} \}$ in equation \eqref{metrica2} as follows:

\begin{equation}
\bar{\mu} (r) \to  \mu(r)= \bar{\mu} (r) - \alpha g(r). \label{deformacionradial}
\end{equation}

This is known as {\it Minimal Geometric Deformation} \cite{Ovalle:2017fgl}.

Replacing equation \eqref{deformacionradial} into \eqref{EqMotion1}, we use the binomial theorem as follows in the left side of this equation \eqref{EqMotion1}:
\begin{align}
   & \bigg [ \bigg (\frac{1-\bar{\mu}(r)}{r^2} \pm \frac{1}{l^2} \bigg) + \bigg ( \alpha \frac{g}{r^2} \bigg) \bigg ]^{n}
    =\bigg (\frac{1-\bar{\mu}(r)}{r^2} \pm \frac{1}{l^2} \bigg)^n+
\nonumber \\
    & \bigg ( \begin{array}{c}  n\\ 1  \end{array} \bigg)
    \bigg (\frac{1-\bar{\mu}(r)}{r^2} \pm \frac{1}{l^2} \bigg)^{n-1}\bigg ( \frac{g}{r^2}   \bigg) \alpha +...+ \bigg ( \begin{array}{c}  n\\ n-1  \end{array} \bigg)
    \bigg (\frac{1-\bar{\mu}(r)}{r^2} \pm \frac{1}{l^2} \bigg)
\bigg ( \frac{g}{r^2}   \bigg)^{n-1} \alpha^{n-1}+ \nonumber \\ &\bigg ( \frac{g}{r^2}   \bigg)^{n} \alpha^{n}
\end{align}

Thus, the system splits into the following sets of equations:
\begin{itemize}
    \item The standard LUV equations for a seed solution (with $\alpha=0$):
   \begin{align} \label{EqMotion2}
\frac{d}{dr} \bigg (  r^{d-1} \bigg [ \frac{1-\bar{\mu}(r)}{r^2} \pm \frac{1}{l^2}  \bigg ]^{n}  \bigg ) = r^{d-2} \bar{\rho}
\end{align}

   and the conservation equation given by:
    \begin{equation} \label{conservacioncerocaso}
     \bar{p}_r'+ \frac{d-2}{r}(\bar{p}_r-\bar{p}_\theta )=0      \end{equation}
    \item The terms of order $\alpha$ give rise to the quasi-LUV equations of order $\alpha^1$,
    \begin{equation} \label{EqMotionUno}
     \frac{d}{dr} \bigg (   n \cdot r^{d-1} \bigg (\frac{1-\bar{\mu}(r)}{r^2} \pm \frac{1}{l^2} \bigg)^{n-1}\bigg ( \frac{g}{r^2}   \bigg) \bigg )
    =- r^{d-2} (\theta_1)^0_0
    \end{equation}
        and the conservation equation given by:
    \begin{equation} \label{conservacionunocaso}
        \big ((\theta_1)^1_1 \big )' +\frac{d-2}{r} \big ((\theta_1)^1_1-(\theta_1)^2_2 \big )=0
    \end{equation}
Thus, following the iteration, it is possible to obtain the quasi LUV equations of order $\alpha^2$, $\alpha^3$...$\alpha^{n-3}$, $\alpha^{n-2}$.

\item The terms of order $\alpha^{n-1}$ give rise to the following  quasi LUV equations of order $\alpha^{n-1}$:
\begin{equation} \label{EqMotionn1}
 \frac{d}{dr} \bigg (   n \cdot r^{d-1} 
    \bigg (\frac{1-\bar{\mu}(r)}{r^2} \pm \frac{1}{l^2} \bigg)
\bigg ( \frac{g}{r^2}   \bigg)^{n-1} \bigg )=- (\theta_{n-1})^0_0
\end{equation}
and the conservation equation given by:
\begin{equation} \label{conservacionn1caso}
    \big ((\theta_{n-1})^1_1 \big )' +\frac{d-2}{r} \big ((\theta_{n-1})^1_1-(\theta_{n-1})^2_2 \big )=0
\end{equation}
\item The terms of order $\alpha^{n}$ give rise to the quasi-LUV equations of order $\alpha^{n}$:
\begin{equation} \label{EqMotionn}
     \frac{d}{dr} \bigg ( r^{d-1}   \bigg ( \frac{g}{r^2}   \bigg)^{n} \bigg )=-r^{d-2} (\theta_n)^0_0
\end{equation}
and the conservation equation given by:
\begin{equation} \label{conservacionncaso}
    \big ((\theta_n)^1_1 \big )' +\frac{d-2}{r} \big ((\theta_n)^1_1-(\theta_n)^2_2 \big )=0 .
\end{equation}
\end{itemize}

It is worth stressing that each quasi LUV equation cannot be formally identified as the spherically symmetric LUV equations of motion for $n>1$, because the left side of each quasi LUV equation do not have the standard form given by the equation \eqref{EqMotion1}. Furthermore, the Bianchi identities are not satisfied for each quasi LUV equation. For $n=1$ the quasi Einstein equations can be transformed into the standard Einstein equations after a convenient redefinition of the energy momentum  tensor \cite{Ovalle:2017fgl}, however, the method of the reference \cite{Ovalle:2017fgl} has been widely used to find new solutions without using this mentioned redefinition in several works.

Despite the above mentioned, our imposed way for solving the original system \eqref{EqMotion1},  \eqref{conservacionImpuesta1} and \eqref{conservacionImpuesta2}, based in the decoupling of sources by means of the standard and quasi LUV equations, ensures us to solve successfully this original system.

Furthermore, under our assumptions, each conservation equation \eqref{conservacioncerocaso}, \eqref{conservacionunocaso},$...$  \eqref{conservacionn1caso}, \eqref{conservacionncaso} is separately conserved, and thus, there is no exchange of energy momentum between the seed fluid and each sector $(\theta_i)_{A B}$. So, in our gravitational decoupling method there is only purely gravitational interaction.

\section{Conserved charge in LUV by Gravitational Decoupling}

In this section we test the effect of the addition of extra sources in the energy momentum tensor on the computation of the energy of the black hole solutions obtained by the GD algorithm for Lovelock with Unique Vacuum theory.

Following \cite{Aoki:2020prb}, the energy is defined as:
\begin{equation} \label{CargaConservada}
    E = \int d^{d-1}x \sqrt{-g} J^0 = \int d^{d-1}x \sqrt{-g} T^0_\nu \xi^\nu,
\end{equation}
where $\xi^\nu$ is a Killing vector and $d$ is the number of dimensions. 

In our case the energy momentum tensor with $n$ extra sources is given by the equation \eqref{EM1}. Also following \cite{Aoki:2020prb}, we choose $\xi^\mu=-\delta^\mu_0$. For the line element \eqref{metrica2} the energy is given by:
\begin{equation}
 \displaystyle   {E}=-\Omega_{d-2} \int^\infty_0 r^{d-2}  T^0_0 dr= \Omega_{d-2} \int^\infty_0  r^{d-2} {\rho}(r) dr,
\end{equation}
where the energy density is given by the equation \eqref{densidadefectiva}. Given that, due to the application of the GD algorithm, the energy density is decoupled in one seed energy density and $n$ extra sources, the energy splits in the following sectors:

\subsection{Energy of the seed solution:}

This corresponds to the energy of the seed solution and is computed with the $\bar{T}^0_0$ component of the seed energy momentum 
\begin{equation} \label{EnergiaSeed}
 \displaystyle   \bar{E}=-\Omega_{d-2} \int^\infty_0 r^{d-2}  \bar{T}^0_0 dr= \Omega_{d-2} \int^\infty_0  r^{d-2} \bar{\rho}(r) dr
\end{equation}

In this work we assume that the seed solution represents a black hole solution by itself, so, the energy of the seed solution will be assumed of positive sign, $\bar{E}>0$. 

\subsection{Energy of each quasi-LUV sector:}

Corresponds to the contribution of each quasi-LUV sector of order $\alpha^i$, with $i=1..n$, whose source is given by the component $(\Theta_i)^0_0$ on the energy density \eqref{densidadefectiva}:
\begin{equation} \label{ContribucionQuasi}
 \displaystyle   E_i= - \Omega_{d-2} \int^\infty_0  r^{d-2} (\Theta_i)^0_0 dr
\end{equation}

\subsection{The total energy:} 
Corresponds to the energy of the Minimal Geometric Deformed solution \eqref{deformacionradial}. This energy corresponds to the sum of the energy of the seed solution plus the contribution of each quasi-LUV sector of order $\alpha^i$. Thus, the total energy is:

\begin{equation} \label{EnergiaTotal}
    E= \bar{E}+ \alpha E_1 + \alpha^2 E_2+...+\alpha^{n-1} E_{n-1}+\alpha^n E_n
\end{equation}

The value (positive or negative) of the energy of each quasi-LUV sector depends of the form of each source $(\Theta_i)^0_0$. So, assuming that the energy of the seed solution is positive, both the form of the sources as the $\alpha$'s constants must be such that the total energy is positive. 

So, one consequence of the application of the gravitational decoupling algorithm to the seed solution is that the total mass increases or decreases due to the contribution of energy of each quasi-LUV sector. The number of extra contributions coincides with the power $n$ of the Riemann tensor in the action. So, the difference between the energy of the Minimally Geometric Deformed solution ($E$) and the energy of the seed solution ($\bar{E}$) is:

\begin{equation} \label{DeltaE}
    \Delta E =E-\bar{E}= \alpha E_1 + \alpha^2 E_2+...+\alpha^{n-1} E_{n-1}+\alpha^n E_n,
\end{equation}
which represents the quantity in which the energy increases or decreases due to the application of the GD algorithm. So, this quantity corresponds to the sum of each contribution of energy of each quasi-LUV sector multiplied by the $\alpha^i$ constant, where $i$ represents each power of the Riemann tensor in the Lovelock action. 

So, the quantity $\Delta E$ must be such that $\bar{E}+ \Delta E>0$.

It is worth to mention that, in the case that both $E_i$ and $\alpha^i$ are positive, applying GD to the LUV, the energy increases more, with the higher value of $n$ chosen. This latter is because the number of extra sources is determined by the value of $n$, {\it i.e} depend on the power of the Riemann tensor in the action.

\section{The New Solution for LUV by gravitational decoupling} \label{seccion5}

To find a new solution we propose the following strategy:

\begin{enumerate}
    \item Pick up a seed solution $\{ \bar{\mu},\bar{\rho}\}$ of the Standard LUV equation \eqref{EqMotion2} .
    
    In this connection we choose the following generalization of the vacuum and asymptotically AdS solution of reference \cite{Crisostomo:2000bb}:
    \begin{equation} \label{solucionTotal}
    \bar{\mu}=1+\frac{r^2}{l^2} - \left ( \frac{2M \cdot H(r)}{\Omega_{d-2} \cdot r^{d-2n-1} } \right)^{1/n}
    \end{equation}
with $M>0$ and where $H(r)$ corresponds to the Heaviside function. The cosmological constant \eqref{ConstanteCosmologica} is negative, and $\Omega_{d-2}$ is the area of a unitary $(d-2)$ sphere. The equation \eqref{solucionTotal} corresponds to the seed solution. The inclusion of the central Heaviside function (which is not present in \cite{Crisostomo:2000bb}) will allow us to compute the mass of the seed solution as we will see bellow. 

Replacing the equation \eqref{solucionTotal} into the standard LUV equation of motion \eqref{EqMotion2} for negative cosmological constant, the seed energy density is:
\begin{equation} \label{SeedEnergyDensity}
\frac{2M}{\Omega_{d-2}  r^{d-2}} \cdot \delta(r) = \bar{\rho}
\end{equation}

The inclusion of the Heaviside function on the singular term of the Schwarzschild AdS solution for the Einstein Hilbert gravity was proposed in reference \cite{Aoki:2020prb}. It is worth to mention that the inclusion of this function on the seed solution (43) $\bar{\mu}$ generates a Dirac delta distribution in the components of the LUV equation of motion that include $\bar{\mu}$ (only the quasi-LUV equations of order $\alpha^n$ do not have presence of $\bar{\mu}$), which must be compensated with a Dirac delta distribution on these components of the energy momentum tensor. The singularity of this central step function is protected by the horizon radius.

\item We solve the quasi LUV equation of order $\alpha^n$. For this, we write $g(r)$ as:
\begin{equation}
    g(r)= \left (\frac{2m(r)}{\Omega_{d-2} \cdot r^{d-2n-1}} \right)^{1/n}.
\end{equation}
So, from equation \eqref{EqMotionn}, we obtain that:
\begin{equation} \label{funciondemasa}
    m(r)=- \Omega_{d-2} \int r^{d-2} (\Theta_n)^0_0 dr .
\end{equation}

So, from equation \eqref{deformacionradial}, the solution is:

\begin{equation} \label{Solucion1}
    \mu(r)=1+\frac{r^2}{l^2} - \left ( \frac{2M \cdot H(r)}{\Omega_{d-2} \cdot r^{d-2n-1} } \right)^{1/n} - \alpha \left (\frac{2m(r)}{\Omega_{d-2} \cdot r^{d-2n-1}} \right)^{1/n}.
\end{equation}

The source $(\Theta_n)^2_2$ is computed by the Quasi LUV equation of order $\alpha^n$, equation \eqref{conservacionncaso}.

\item Once we know the functions $\bar{\mu} (r)$ and $g(r)$, we compute directly the sources $(\theta_i)^0_0$ in the quasi LUV equations of order $\alpha^1...\alpha^{n-1}$. The sources $(\theta_i)^2_2$ are computed by the corresponding conservation equation of each quasi LUV equation.
\end{enumerate} 

Thus, the equation \eqref{Solucion1} represents a new type of black hole solution by gravitational decoupling for the LUV theory. It is worth to mention that this solution has a central singularity. So, this singularity must be protected by the black hole horizon. For the limit $\alpha \to 0$ and for $r>0$ our solution coincides with reference \cite{Crisostomo:2000bb}. 

On the other hand, it is well known that both the Pure Lovelock vacuum solution \cite{Cai:2006pq} and the Einstein Gauss Bonnet solution \cite{Torii:2005nh}, apart from the central singularity, have a {\it singularity of curvature} (in \cite{Torii:2005nh} is called branch singularity) for negative cosmological constant. A point of singularity of curvature is where the Ricci scalar diverges but the function $f(r)$ does not diverge. Adding one extra source to the LUV vacuum solutions \cite{Crisostomo:2000bb}, without the gravitational decoupling method, also appears a singularity of curvature. Thus, one advantage of our extension of the GD method is that this latter leads to solutions free of singularities of curvature.

It is worth to mention that the main difference between the solution of reference \cite{Crisostomo:2000bb} and our solution \eqref{Solucion1} is that the energy of the seed solution \cite{Crisostomo:2000bb} is given by the equation \eqref{EnergiaSeed}, whereas the energy of our solution corresponds to the sum of the energy of the seed solution plus the contribution of each quasi LUV sector, {\it i.e.} the energy of our solution is given by the equation \eqref{EnergiaTotal}, where the number of contributions depends on the power $n$ of the Riemann tensor in the action. So, due to the application of the GD algorithm the energy increases or decreases in a quantity equal to equation \eqref{DeltaE}. Furthermore, as we will see below, the thermodynamics analysis of our solution also provides new characteristics respect to other solutions previously studied in the literature. 

\subsection{Toy Model}
Still, we have not proposed any form for $(\Theta_n)^0_0$. In this connection we will use the generalization of Hayward density of reference \cite{Aros:2019quj}
\begin{equation}
  (\Theta_n)^0_0 =- \frac{d-1}{\Omega_{d-2}} \frac{L \tilde{M}^2}{(L\tilde{M}+r^{d-1})^2} .
\end{equation}
where $\tilde{M}>0$ represents the mass in the generalization of the Hayward metric \cite{Aros:2019quj} and $L$ is a constant of integration.
Replacing into equation \eqref{funciondemasa}:
\begin{equation}
    m(r)= C -\frac{L\tilde{M}^2}{L\tilde{M}+r^{d-1}}
\end{equation}
Choosing arbitrarily the integration constant $C=\tilde{M}$:
\begin{equation}
    m(r)=\frac{\tilde{M}r^{d-1}}{L\tilde{M}+r^{d-1}}.
\end{equation}
Replacing in equation \eqref{Solucion1}:
\begin{equation} \label{Solucion2}
    \mu(r)=1+\frac{r^2}{l^2} - \left ( \frac{2M \cdot H(r)}{\Omega_{d-2} \cdot r^{d-2n-1} } \right)^{1/n} - \alpha \left ( \frac{\tilde{M}r^{2n}}{L \tilde{M}+r^{d-1}} \right)^{1/n}.
\end{equation}
where $d-2n-1>0$ to ensure the behavior asymptotically AdS

\section{Example: Computing the energy in the simplest case}

As was above mentioned, for $n=1$ the Einstein Hilbert action is recovered. So, in this section we will compute the energy for the simplest case of the Lovelock with Unique Vacuum, wich corresponds to the Einstein Hilbert theory. 

Using the toy model, for $n=1$, the only temporal component of the extra source is given by:

\begin{equation} \label{TemporalExtra}
  (\Theta_1)^0_0 = -\frac{d-1}{\Omega_{d-2}} \frac{L \tilde{M}^2}{(L\tilde{M}+r^{d-1})^2},
\end{equation}
and the corresponding solution is:
\begin{equation} \label{Solucion2}
    \mu(r)=1+\frac{r^2}{l^2} -  \frac{2M \cdot H(r)}{\Omega_{d-2} \cdot r^{d-3} } - \alpha \cdot \frac{\tilde{M}r^{2}}{L \tilde{M}+r^{d-1}} .
\end{equation}

\subsection{Energy of the seed solution}
The energy of the seed solution is obtained by inserting the energy density of the seed solution, equation \eqref{SeedEnergyDensity}, into the equation \eqref{EnergiaSeed}:
\begin{equation}
    \bar{E}=M
\end{equation}

\subsection{Energy of the quasi-LUV sector of order $\alpha^1$}

It is worth to mention that for $n=1$, where the Lovelock Unique Vacuum and Einstein Hilbert theory coincide, the quasi-LUV sector is similar to the quasi-Einstein sector of the reference \cite{Ovalle:2017fgl}. So, the contribution of the temporal component of the extra source is obtained by replacing the equation \eqref{TemporalExtra} into the equation \eqref{ContribucionQuasi}:
\begin{equation}
    E_1=\tilde{M}.
\end{equation}

\subsection{Total Energy}
The total energy corresponds to the sum of the energy of the seed solution plus the contribution of the extra source multiplied by the constant $\alpha$:
\begin{equation}
    E=M + \alpha \cdot \tilde{M}.
\end{equation}
So, for $\alpha>0$, in our example, due to the application of the GD algorithm, the energy  increases proportionally to the constant $\alpha$. This quantity is: 
\begin{equation}
    \Delta E =E-\bar{E}= \alpha \cdot \tilde{M}. 
\end{equation}

\section{About the Thermodynamics of the new solution in LUV}
As example, in this work, we will analyze the solution \eqref{Solucion2} for $n=2$ with $d=7$ and for $n=3$ with $d=8$. It is direct to check that the behavior of the solutions showed is generic for other values of $d$. 

\begin{figure} 
   \begin{center}
      \includegraphics[width=0.49 \textwidth]{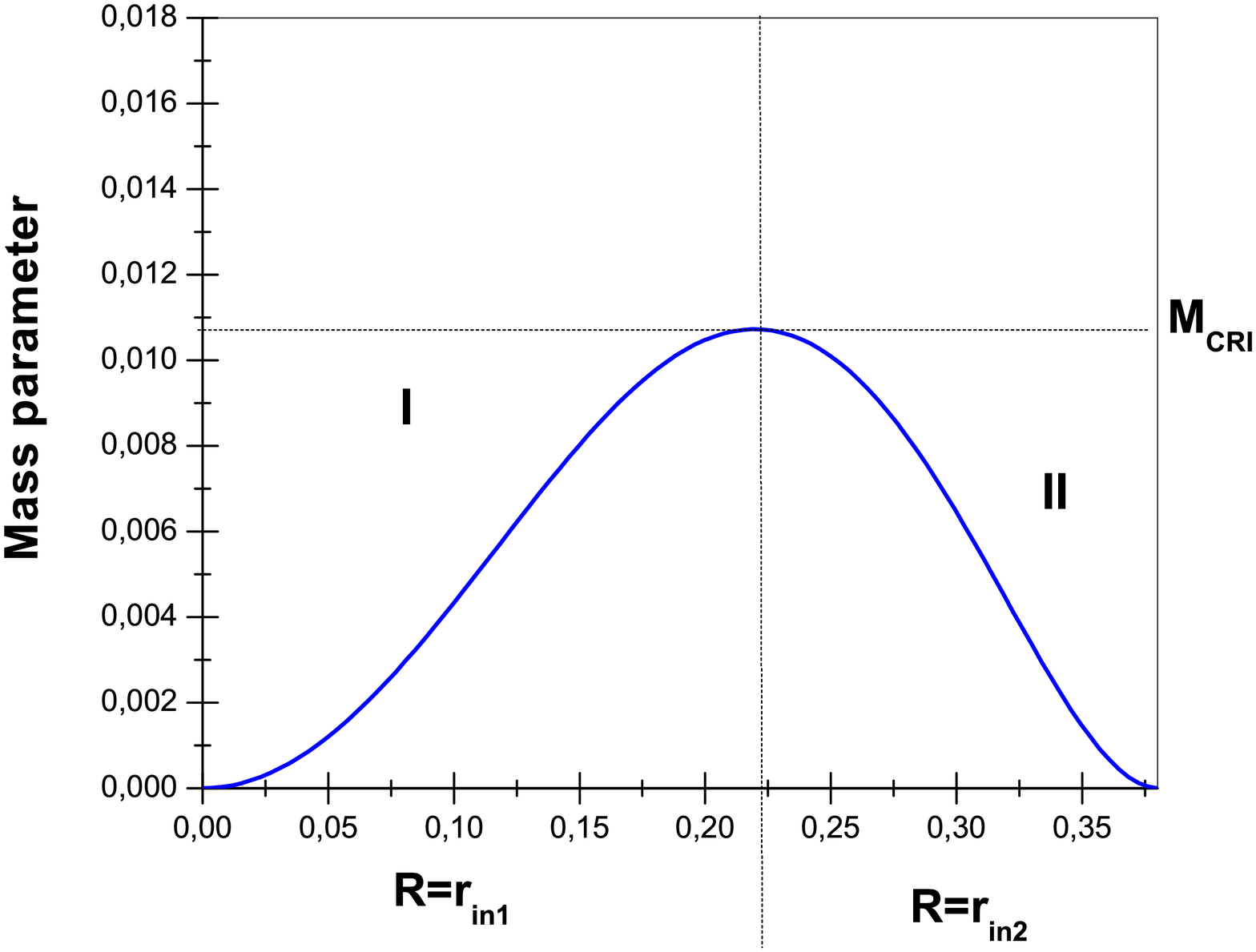} \ 
      \includegraphics[width=0.49 \textwidth]{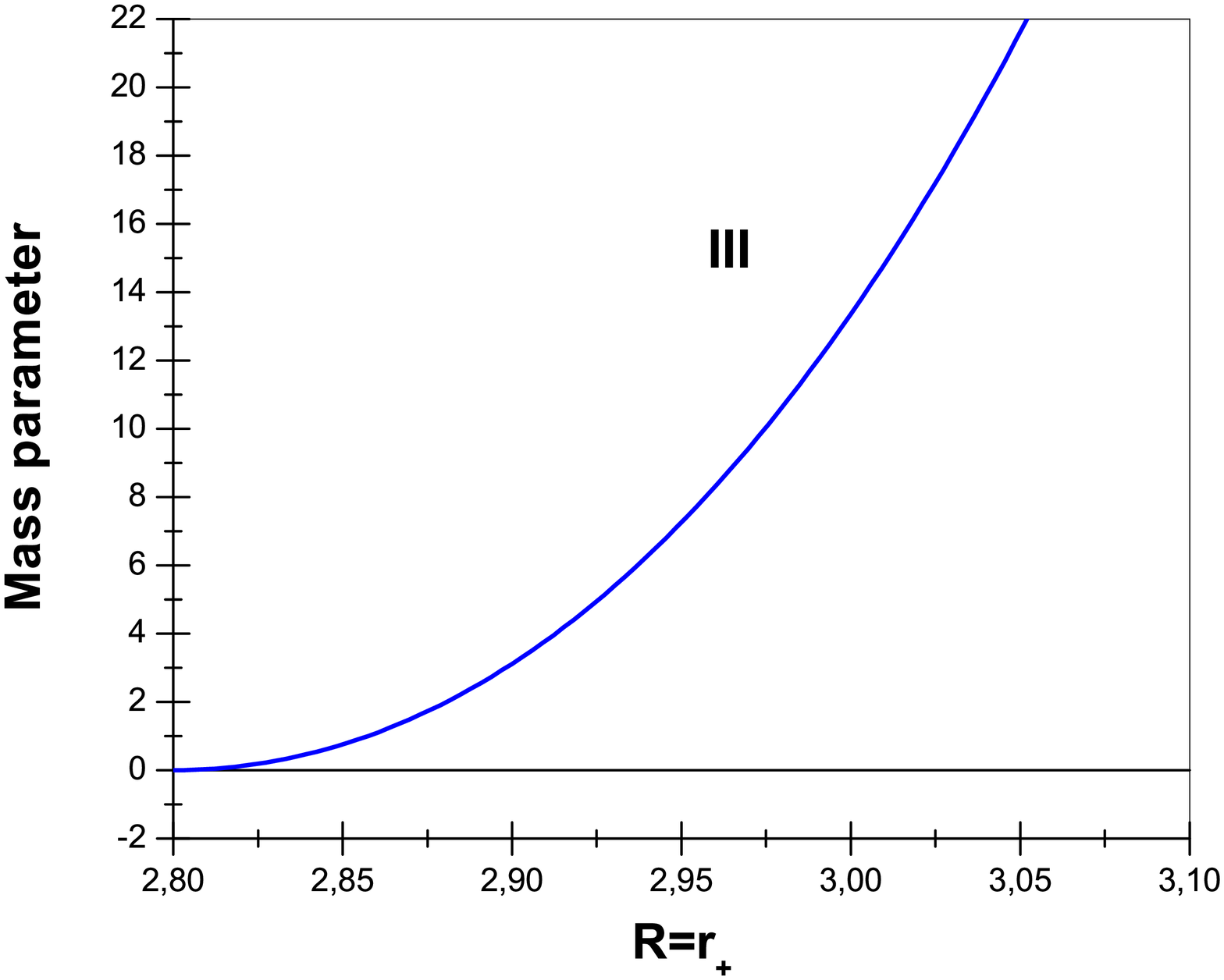} \ 
      \includegraphics[ width=0.49 \textwidth]{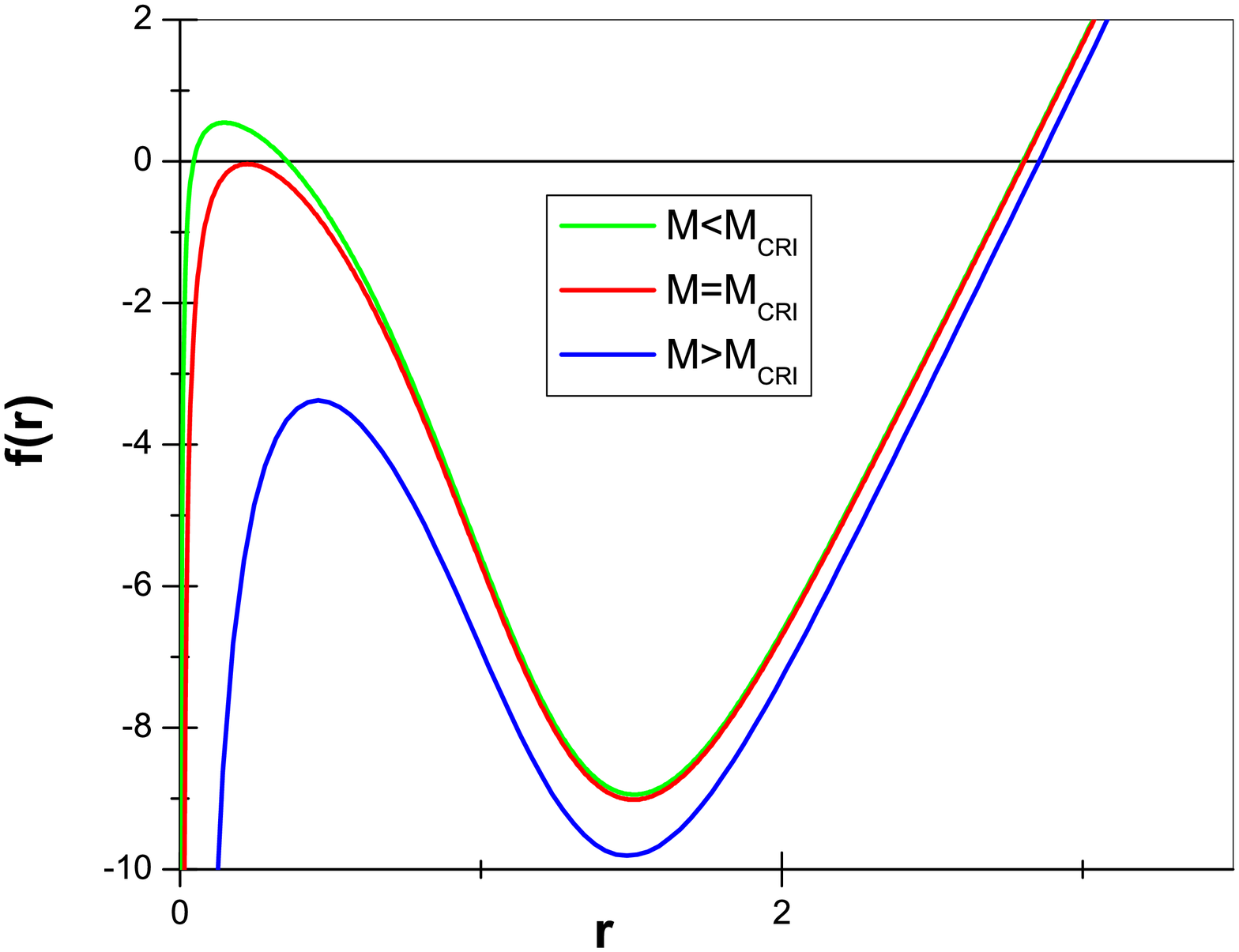}
      \caption{\label{FigTermodinamica} \textbf{first panel:} first part of Mass Parameter. \textbf{second panel:} second part of Mass Parameter. \textbf{Third Panel:} f(r).} 
   \end{center}
\end{figure}

\begin{figure} 
   \begin{center}
      \includegraphics[width=0.49 \textwidth]{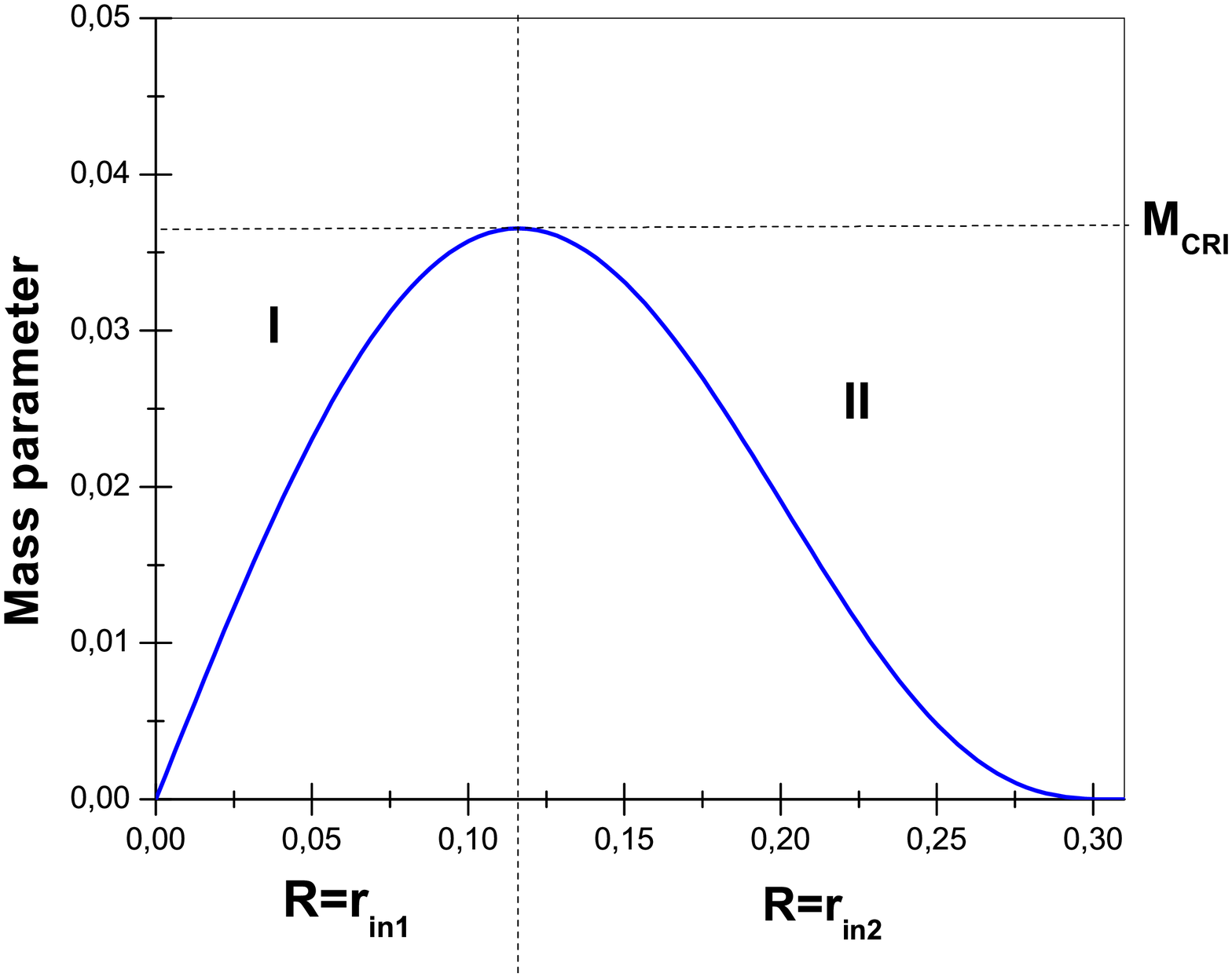} \ 
      \includegraphics[width=0.49 \textwidth]{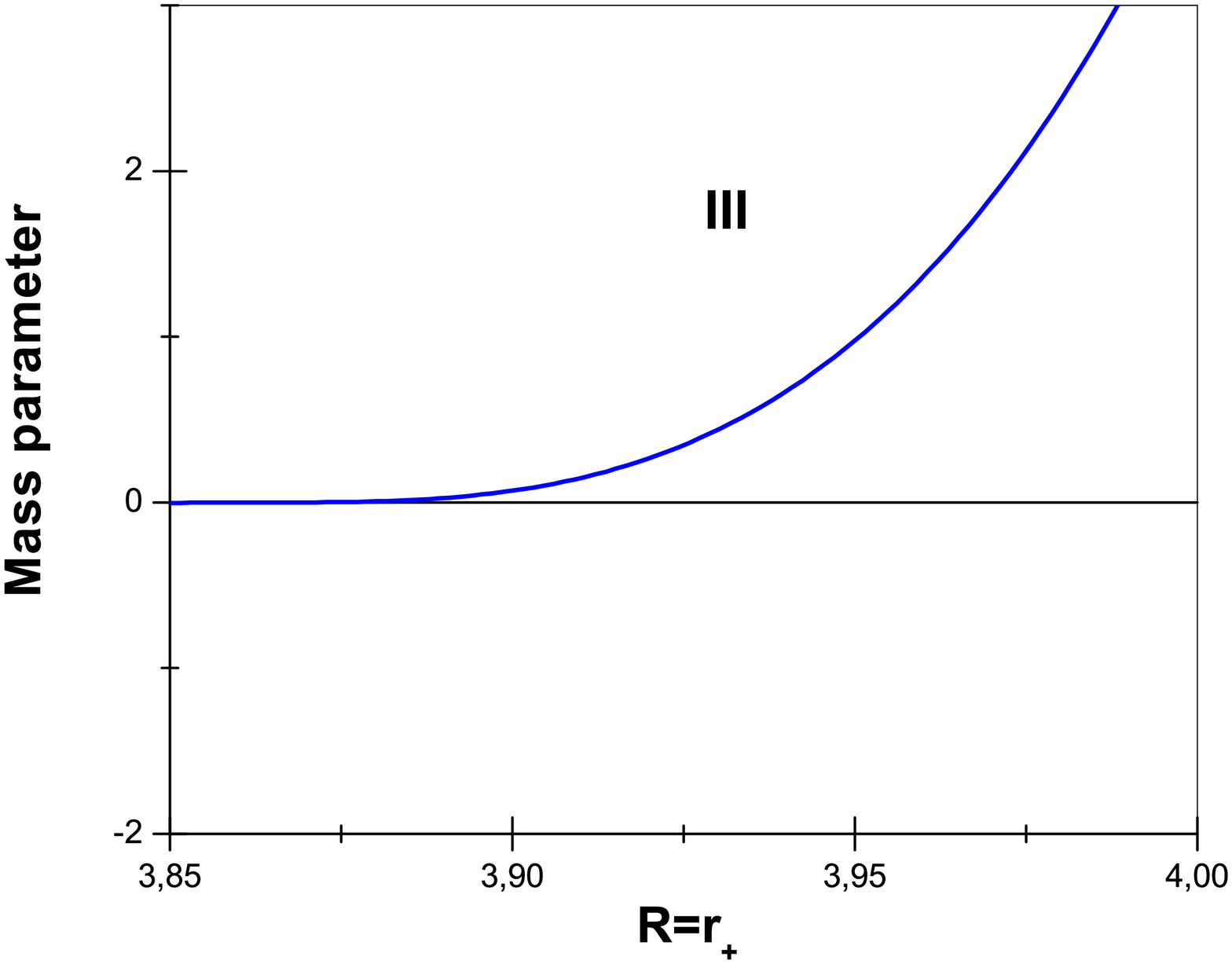} \ 
      \includegraphics[ width=0.49 \textwidth]{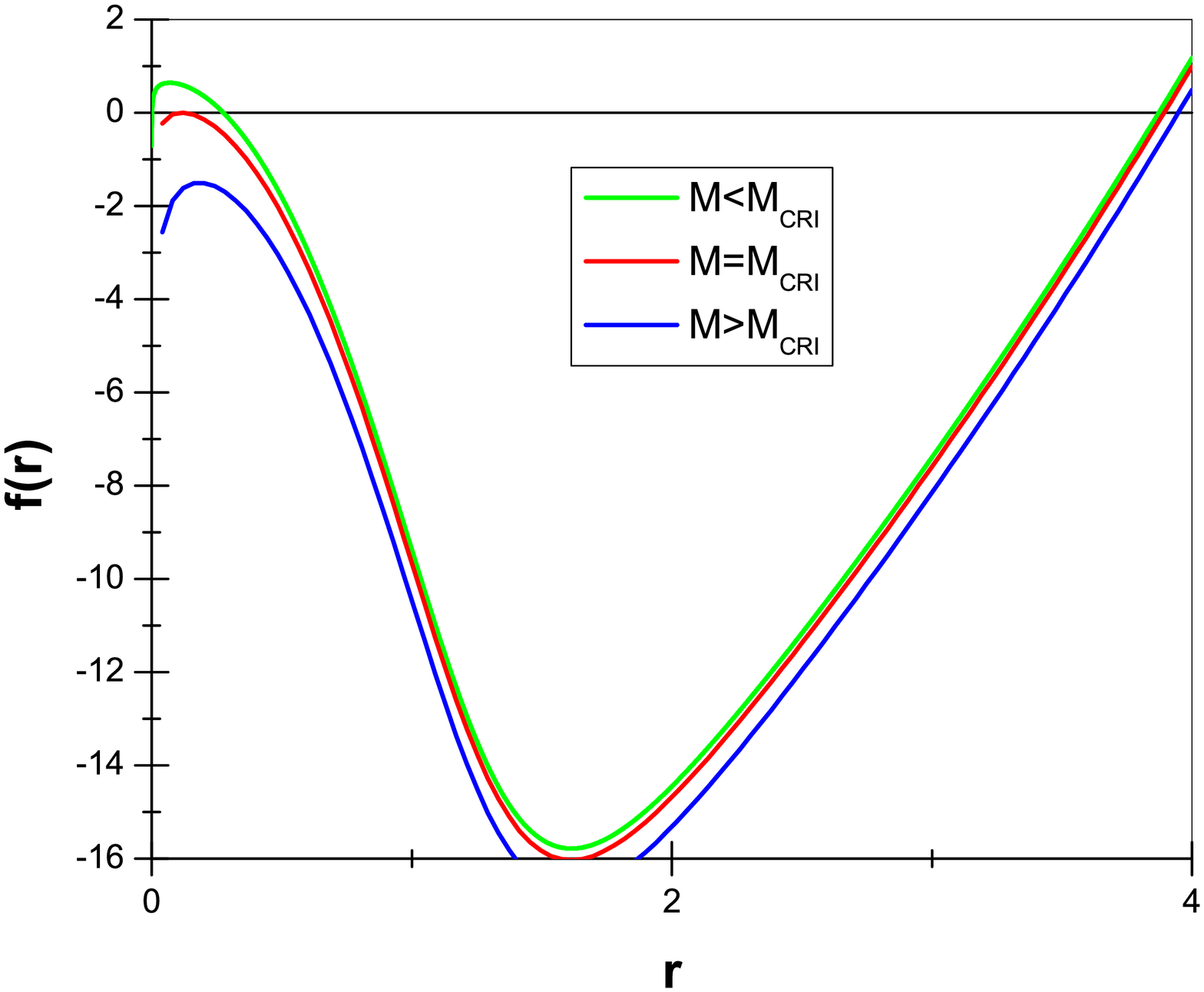}
      \caption{\label{FigTermodinamica1} \textbf{first panel:} first part of Mass Parameter. \textbf{second panel:} second part of Mass Parameter. \textbf{Third Panel:} f(r).} 
   \end{center}
\end{figure}

\subsection{Mass parameter and horizon structure}

The horizon structure is displayed for $n=2$ with $d=7$ and for $n=3$ with $d=8$ in figures \ref{FigTermodinamica} and \ref{FigTermodinamica1}, respectively. It is direct to check that these behaviors are generic for other values of $d$.
For an asymptotically AdS space time, the black hole horizon corresponds to the  largest root of
the equation $f(r=R)=0$ \cite{Aranguiz:2015voa}. This differs from the space time with positive cosmological constant, where the radial coordinate is time--like for large enough  values of $r$ and so it is formed a cosmological horizon \cite{Dolan:2012jh}.

We can see, in the first and second panel of the figures  \ref{FigTermodinamica} and \ref{FigTermodinamica1}, that for a value of the mass parameter $M<M_{CRI}$, the black hole has three horizons: the inner horizon one, the inner horizon two and the black hole horizon. For $M=M_{CRI}$ both inner horizons coincide. For $M>M_{CRI}$ the solution only has the black hole horizon. In the third panel we see the behavior of the function $f(r)$, which can have up to three horizons. The inner horizons are protected for the black hole horizon, so, a priory the inner horizons do not have physical interest.

The existence of two inner horizons is a proper characteristic of our solution. In the vacuum solution of LUV of references \cite{Crisostomo:2000bb,Aros:2000ij} there are not inner horizons. In the regular black hole of LUV of reference \cite{Aros:2019quj} there is only one inner horizon.

\begin{figure} 
   \begin{center}
      \includegraphics[width=0.49 \textwidth]{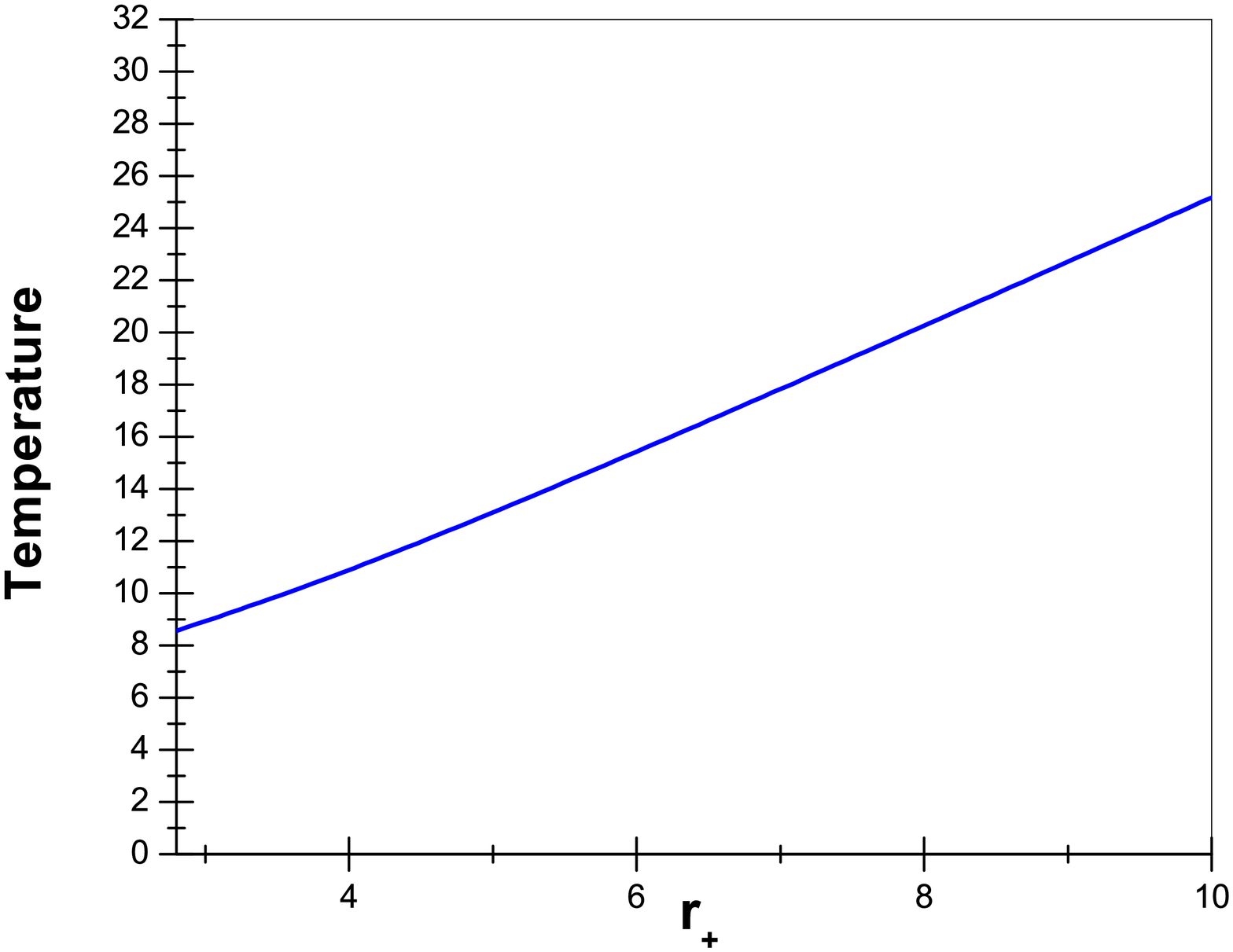} \ 
      \includegraphics[width=0.49 \textwidth]{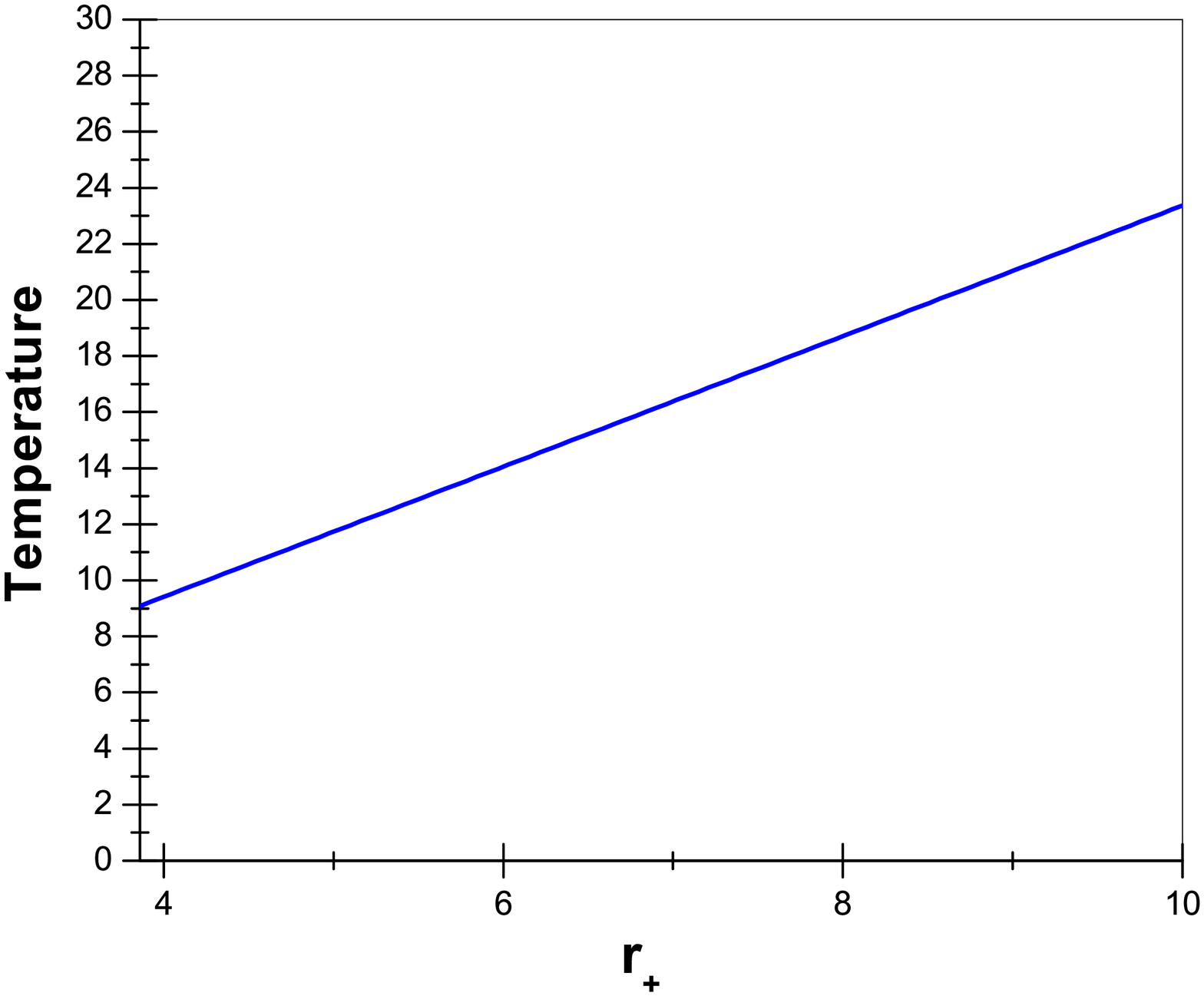} \  \caption{\label{FigTermodinamica4} \textbf{first panel:} Temperature for $n=2$ and $d=7$. \textbf{second panel:} Temperature for $n=3$ and $d=8$. } 
   \end{center}
\end{figure}

\subsection{Temperature and stability}
In the figure \ref{FigTermodinamica4} we show the behavior of the temperature for $n=2$ with $d=7$ and for $n=3$ with $d=8$, respectively. In both cases the temperature is an increasing function of the black hole horizon. This behavior differs from the vacuum solution LUV of reference \cite{Crisostomo:2000bb} where the temperature changes from a decreasing function to an increasing function, and thus, there is a phase transition. Our temperature also differs from the regular black hole solution of LUV of reference \cite{Aros:2019quj}, where there are two phase transitions. In our case, due that $dT/dr_+$ is always positive, the specific heat $C=(dM/dr_+)(dT/dr_+)^{-1}$ is always positive. Thus our solution is always stable and does not have phase transitions.

\subsection{About the first law of thermodynamics}

It is well known that the first law of thermodynamics
is modified due to the presence of matter fields
in the energy momentum tensor \cite{Ma:2014qma}. The inclusion of matter fields forces to redefine the energy term in order to obtain correct values of temperature and entropy.

Recently in reference \cite{Estrada:2020tbz} was proposed a new version of the first law of thermodynamics for the Einstein Hilbert theory in $(2+1)$ dimensions, where a local definition of the variation of energy is defined, so, the computed values of entropy and temperature are correct. 

Following \cite{Estrada:2020tbz} we use the conditions $\mu(r_+,M, \tilde{M})=0$ and $\delta \mu(r_+,M,\tilde{M})=0$, which can be viewed as constraints on the evolution along the space parameters \cite{Aros:2019auf,Zeng:2019huf}. In a future work, could
be studied the case where the constant $L,l$ and $\alpha$ are  thermodynamics parameters.
Thus, from the last condition
\begin{equation} \label{Restriccion}
    0 = \frac{\partial \mu}{\partial r_+} dr_+ + \frac{\partial \mu}{\partial M} dM + \frac{\partial \mu}{\partial \tilde{M}} d\tilde{M}.
\end{equation}

For simplicity, we rewrite the solution \eqref{Solucion1} at the horizon $r_+>0$ (so $H(r=r_+)=1$) as

\begin{equation} \label{SolucionReesc}
    \mu(r_+)=0=1+\frac{r_+^2}{l^2} - \left ( \frac{2 \bar{m}(r_+,M,\tilde{M})}{\Omega_{d-2} \cdot r_+^{d-2n-1} } \right)^{1/n},
\end{equation}
where
\begin{equation}
  \left (  \bar{m}(r_+,M,\tilde{M}) \right )^{1/n}=M^{1/n}-\alpha \cdot \left ( m(r_+,\tilde{M}) \right )^{1/n}.
\end{equation}

We will write the following derivatives as:
\begin{align}
\frac{\partial \mu}{\partial M}&= \frac{\partial \mu}{\partial \bar{m}} \frac{\partial \bar{m}}{\partial M}  =- \left ( \frac{2 \bar{m}}{\Omega_{d-2} \cdot r_+^{d-2n-1} } \right)^{1/n} \cdot \frac{1}{n \cdot \bar{m}} \cdot \frac{\partial \bar{m}}{\partial M} \label{parte1} \\
\frac{\partial \mu}{\partial \tilde{M}}&= \frac{\partial \mu}{\partial \bar{m}} \frac{\partial \bar{m}}{\partial \tilde{M}}  =- \left ( \frac{2 \bar{m}}{\Omega_{d-2} \cdot r_+^{d-2n-1} } \right)^{1/n} \cdot \frac{1}{n \cdot \bar{m}} \cdot \frac{\partial \bar{m}}{\partial \tilde{M}} \label{parte2}
\end{align}

Manipulating equations \eqref{SolucionReesc}, \eqref{parte1} and \eqref{parte2}, equation \eqref{Restriccion} is rewritten as:
\begin{equation}
    \frac{\partial \bar{m}}{\partial M}dM+\frac{\partial \bar{m}}{\partial \tilde{M}} d \tilde{M}=  \left ( \frac{1}{4\pi} f'\big |_{r=r_+}  \right )  \left (\Omega_{d-2} 2 \pi n r_+^{d-2n-1} \left (1+ \frac{r_+^2}{l^2} \right )^{n-1} dr_+\right ).
\end{equation}

The above equation can be rewritten as:
\begin{equation}
    du=TdS,
\end{equation}
where we can identify the temperature and entropy terms as:
\begin{align}
    T=&\frac{1}{4\pi} f'\big |_{r=r_+} \\
   dS=&\Omega_{d-2} 2 \pi n r_+^{d-2n-1} \left (1+ \frac{r_+^2}{l^2} \right )^{n-1} dr_+
\end{align}
values that are correct. In the references \cite{Crisostomo:2000bb,Aros:2000ij} was shown that the entropy for LUV has the form $dS \sim  n r_+^{d-2n-1} \left (1+ \dfrac{r_+^2}{l^2} \right )^{n-1} dr_+$. Furthermore the term
\begin{equation}
    du=\frac{\partial \bar{m}}{\partial M}dM+\frac{\partial \bar{m}}{\partial \tilde{M}} d \tilde{M}
\end{equation}
corresponds to a local definition of the variation of energy at the horizon. For $n>1$ is not possible to split the first law in the standard and quasi sectors, due to the dependence on $n$ of the LUV equations. However, we have provided a new form for the first law of thermodynamics with presence of matter fields in the energy momentum tensor for LUV gravity, which allows to obtain the correct values of temperature and entropy.

\section{Conclusion and Discussion}
We have provided an extension of the Gravitational Decoupling algorithm for LUV. The application of our extension provide a simple way to solve the original system of equations of motion. In this extension, the number of extra sources is determined by the value of $n$, {\it i.e} depend on the power of the Riemann tensor in the action.
In this extension, the equations of motion split in the standard LUV equations of motion and in the quasi LUV of order $\alpha^i$ equations.

Under the assumptions imposed , each conservation equation is conserved, and thus, there is no exchange of energy momentum between the seed fluid and each sector $(\theta_i)_{A B}$. So, in our gravitational decoupling method there is only purely gravitational interaction. 
It is wort to mention that the equation of motion for LUV is only possible for the case where $-g_{tt}=(g_{rr})^{-1}$.

 On the other hand, we have tested the effect of the addition of extra sources in the energy momentum tensor on the computation of the energy of the black hole solutions obtained by the GD algorithm for LUV. For this, we have used the recently definition of conserved charges of reference \cite{Aoki:2020prb}. We have showed that, one consequence of the application of this algorithm is that the energy of the system splits in the {\it energy of the seed solution} and the {\it energy of each quasi-LUV sector}.
 
 The total energy corresponds to the sum of the energy of the seed solution plus the contribution of each quasi-LUV sector of order $\alpha^i$. So, under certain assumptions imposed, the total mass increases or decreases due to the contribution of energy of each quasi-LUV sector. The difference between the energy of the Minimally Geometric Deformed solution and the energy of the seed solution, so--called $\Delta E$, represents the quantity in which the energy varies due to the application of the GD algorithm. This quantity depends on the number of $n$ extra sources, which coincides with the power of the Riemann tensor in the action.
 
 It is worth to mention that the main difference between the solution of reference \cite{Crisostomo:2000bb} and our solution \eqref{Solucion1} is that the energy of the seed solution \cite{Crisostomo:2000bb} is given by the equation \eqref{EnergiaSeed}, whereas the energy of our solution corresponds to the sum of the energy of the seed solution plus the contribution of each quasi LUV sector, {\it i.e.} the energy of our solution is given by the equation \eqref{EnergiaTotal}, where, as was above mentioned, the number of contributions depends on the power $n$ of the Riemann tensor in the action. So, due to the application of the GD algorithm the energy increases or decreases in a quantity equal to equation \eqref{DeltaE}.

On the other hand, the thermodynamics analysis also provides new characteristics of the new solution obtained by GD for LUV:

\begin{itemize}
\item The obtained solution can have up to three horizons depending on the value of the critical mass parameter, namely $M_{CRI}$. For $M<M_{CRI}$, the black hole has three horizons: the inner horizon one, the inner horizon two and the black hole horizon. For $M=M_{CRI}$ both inner horizons coincide. For $M>M_{CRI}$ the solutions only has the black hole horizon.

\item The existence of two internal horizons is a proper characteristic of our solution. In the vacuum solution of LUV of references \cite{Crisostomo:2000bb,Aros:2000ij} there are not inner horizon. In the regular black hole of LUV of reference \cite{Aros:2019quj} there is only one inner horizon.

\item In our solution, the temperature is an increasing function of the black hole horizon. This behavior differs from the vacuum solution LUV of reference \cite{Crisostomo:2000bb} where the temperature changes from a decreasing function to a increasing function, and thus, there is a phase transition. Our temperature also differs from the regular black hole solution of LUV of reference \cite{Aros:2019quj}, where there are two phase transitions. In our case, due that $dT/dr_+$ is always positive, the specific heat $C=(dM/dr_+)(dT/dr_+)^{-1}$ is always positive. Thus our solution is always stable an does not have phase transitions.
\end{itemize}

Since the first law of thermodynamics for black holes is modified by the presence of the matter fields in the energy momentum tensor, we have provided a new version of the first law for LUV, where a local definition of the variation of energy is defined, and, where the entropy and temperature are consistent with the previously known in literature for LUV.

\bibliography{mybib}

\end{document}